\documentstyle[preprint,aps,psfig]{revtex}

\def\Journal#1#2#3#4{{#1} {\bf #2}, #3 (#4)}

\def\NPA{{\em Nucl. Phys.} A}
\def\PLB{{\em Phys. Lett.}  B}
\def\PR{{\em Phys. Rep.} }
\def\PRL{\em Phys. Rev. Lett.}
\def\PRD{{\em Phys. Rev.} D}
\def\ZPC{{\em Z. Phys.} C}
\def\ZPA{{\em Z. Phys.} A}

\begin{document}
\draft
\preprint{Subm. to \ZPC}
\title{Hadron production in relativistic nuclear collisions: thermal 
hadron source or hadronizing quark-gluon plasma?
\footnote[1]{Supported by GSI, BMBF, DFG}}
\author{C.~Spieles, H.~St\"ocker}
\address{Institut f\"ur
Theoretische Physik,  J.~W.~Goethe-Universit\"at,\\
D-60054 Frankfurt am Main, Germany}
\author{C.~Greiner}
\address{Institut f\"ur Theoretische Physik,
J.~Liebig-Universit\"at,\\
D-35392 Giessen, Germany}
\date{\today}
\maketitle
\begin{abstract}
Measured hadron yields from relativistic nuclear
collisions can be equally well understood in two physically distinct 
models, namely a static thermal hadronic source vs.~a time-dependent, 
nonequilibrium hadronization off a quark-gluon plasma droplet.
Due to the time-dependent particle evaporation off
the hadronic surface in the latter approach the hadron ratios change 
(by factors of $\stackrel{\textstyle <}{\approx}5$) in time.
Final particle yields reflect time averages over the actual thermodynamic 
properties of the system at a certain stage of the evolution. 
Calculated hadron, strangelet and
\mbox{(anti-)}cluster yields as well as freeze-out times are presented for
different systems. Due to strangeness distillation the system moves rapidly
out of the $T$, $\mu_q$ plane into the $\mu_s$-sector. Strangeness to baryon
ratios $f_s=1-2$ prevail during a considerable fraction (50\%) of the time
evolution (i.e. $\Lambda$-droplets or even $\Xi^-$-droplets form the system
at the late stage: The possibility of observing this time evolution via HBT
correlations is discussed). The observed hadron ratios require $T_c\approx
160$~MeV and $B^{1/4}\stackrel{\textstyle >}{\approx}200$~MeV.
If the present model is fit to the extrapolated hadron yields,
metastable hypermatter can only be produced with a probability $p< 10^{-8}$
for $A \ge 4$.
\end{abstract}

\section{Hadron ratios from an equilibrated source}
Measurements of strange and anti-strange particles in relativistic nuclear 
collisions
have received much attention recently\cite{qm95,str95}:
(Anti-)strangeness enhancement relative to pp-data has been predicted as 
quark gluon plasma signal, because equilibrium yields of strange hadrons
cannot be achieved during the short collision time due to small $s\bar s$
cross sections\cite{Koc86}. However, several thermal
models have been applied\cite{Cley93,Le95,PBM} to (strange)
particle yields and used to extract the characteristic thermodynamic properties
of the system (a few macroscopic parameters) from chemical equilibrium. 

\begin{figure}[b]
\vspace*{\fill}
\centerline{\psfig{figure=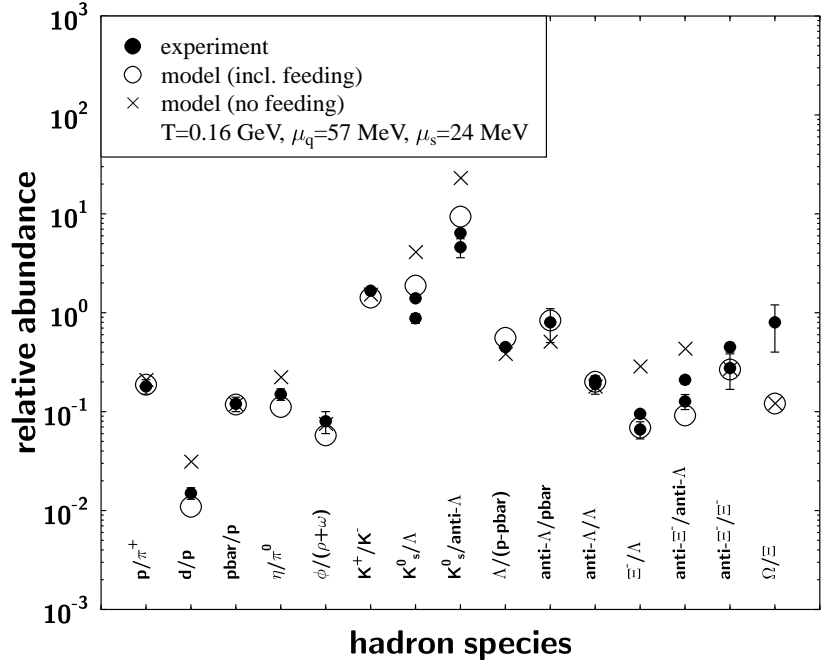,width=12cm}}
\caption{Particle ratios in the static equilibrium scenario with the parameters
(temperature, chemical potentials) as indicated. The crosses denote the
resulting values if contributions due to the decay of higher lying
resonances are ignored. The circles include these effects.
Data from various experiments as compiled in \protect\cite{PBM} 
are also shown.
\label{fs160}}
\vspace*{\fill}
\end{figure}

Fig.~\ref{fs160} shows such a calculation: different measured and 
calculated hadron ratios are shown as resulting from an
equilibrium hadron gas at fixed temperature $T$, quark chemical
potential $\mu_q$ and
strange quark chemical potential $\mu_s$. The constituents of the hadron gas 
are all
well-established hadrons up to masses of 2~GeV (coinciding with the list
given in \cite{Ba96}) taken from\cite{PDB96}. Note that an isospin symmetric
system is assumed (as in the hadronization model): 
\#(protons)=\#(nucleons)/2, \#($\pi^+$)=\#(all pions)/3 etc.
The system is treated as a mixture of relativistic, non-interacting 
 Bose--Einstein and Fermi--Dirac 
gases with Hagedorn's eigenvolume correction.

The parameters $T$ and $\mu_q$ of Fig.~\ref{fs160} are chosen 
in line with
the values given by P.~Braun-Munzinger and J.~Stachel in \cite{PBM}.
Agreement with many
observed ratios
from different experiments at the SPS is achieved (as in \cite{PBM}).
The value of $\mu_s$ has been determined by
the requirement that the total net-strangeness vanishes: $f_s^{init}=0$.
The quantity $f_s$ is the fraction of net-strangeness over net-baryon number
($f_s=(N_s-N_{\bar s})/A$).
Note that the presently calculated strange quark chemical potential 
$\mu_s=24$~MeV is different from the
value $\mu_s=18.6$~MeV given in \cite{PBM}. For the ratios we gain 
roughly the same results,
however, there is a discrepancy of about a factor of 2 in the
$\Lambda/(p-\bar p)$ and $\bar \Lambda/\bar p$ values.

It should be mentioned 
that some of the measured ratios may deviate from the true,
yet to be observed $4\pi$-ratios: the $p/\pi^+$ ratio, which 
is a rather direct
measure of the specific entropy $S/A$ in the fireball, has a value of
$0.18(3)$. However, it is calculated from proton yields in the rapidity
range 2.6 to 2.8 and pion yields in the range 3.2 to 4.8 \cite{NA44}.
The $(\Omega^-+\overline\Omega^-)/(\Xi^-+\overline\Xi^-)$ ratio is measured
for $p_t>1.6$~GeV which excludes a major part of the momentum
distributions. If one takes the value for a common $m_t$-cut ($1.7\pm 0.9$
for $m_t>2.3$~GeV \cite{WA85}) the ratio is compatible with the extracted
temperatures and chemical potentials, since within the restricted
phase space it is just the ratio of the fugacities times the
degeneracies\cite{Le95}, which deviates considerably from the value for the
full phase space.
However, to allow for a consistent comparison to other models
all experimental data are taken from the compilation in \cite{PBM} and we
also give $4\pi$ ratios for both presented models.

Fig.~\ref{fs160} shows the massive effect (factors of
$\stackrel{\textstyle >}{\approx}3$!) of feeding on the
predicted ratios (cf. the discussion in \cite{Hah88}): the crosses denote the
ratios as predicted, if the contributions due to the decay of higher lying
resonances (e.g. $\Delta \rightarrow \pi+p$, $\rho \rightarrow \pi^++\pi^-$
etc.) are ignored. The circles show the ratios as obtained after one 
includes these feeding effects. In some cases
both hadron species change considerably due to feeding, but sometimes such 
that the
ratio does not change much, e.g. $(\bar p/p)^{eq}\approx\frac{\displaystyle 
\bar
p^{eq}+\bar p^*}{\displaystyle p^{eq}+p^*}$, where $\bar p^*$, $p^*$ denote the
yields\cite{Hah88} due
to the decay of resonances (e.g. $\bar \Delta$, $\Delta$).

Collective motion and fits to particle spectra, respectively,
are not considered here. Longitudinal and transverse 
flow velocities have 
already been extracted from particle spectra in combination with a static 
thermal model\cite{St96}. However, it is questionable, whether thermal 
and chemical freeze-out happen simultaneously. Here, we focus 
on a collective, {\em chemical} freeze-out.

\section{The hadronization model}
We now investigate, whether the reasonable agreement of data and model,
as reached with the static equilibrium-plus-feeding model, can also be
achieved with a dynamic model, which includes the formation and expansion of
quark-gluon matter, a first order phase transition into coexistence of quark
and hadron matter, and time-dependent evaporation of hadrons from the
system, as it evolves with time through the phase transition.
This means adopting a model for the hadronization and space-time 
evolution of quark matter droplets given in \cite{CG2}. 
A similar approach was given in \cite{Ba88}, however, in terms of rate
equations for the flavor kinetics. They allow the abundances of the
different constituents of the system to be out of equilibrium, which is in
contrast to our model. The calculated hadron yields in both models are
similar, if the same initial input is used.
%\cite{FriPriv}.

The model \cite{CG2} assumes
 a first order phase transition of the QGP to hadron gas with Gibbs
conditions ($P^{QGP}=P^{HG}$, $T^{QGP}=T^{HG}$, $\mu_q^{QGP}=\mu_q^{HG}$, 
$\mu_s^{QGP}=\mu_s^{HG}$) for coexistence.
The expansion and evaporation of the system takes into account equilibrium,
 as well as nonequilibrium features
of the process:
\begin{enumerate}
\item The plasma sphere is permanently surrounded by a layer of hadron gas,
with which it stays in thermal and chemical equilibrium during the phase
transition (Gibbs conditions).
The strangeness degree of freedom stays in 
chemical equilibrium between the two phases (however, $<s-\bar s>\ne 0$ for
the individual phases). Thus, the hadronic particle production is 
driven by the chemical potentials.  
\item The particle evaporation is incorporated by 
a time-dependent freeze-out of hadrons from the 
hadron phase surrounding the QGP droplet.
During the expansion, the volume increase of the system thus
competes with the decrease due to the evaporative, time-dependent freeze--out.
\end{enumerate}

The global properties, like (decreasing or increasing) $S/A$ and $f_s$ 
 of the remaining two-phase system, then change in time
according to the following differential equations for the baryon number, the
entropy, and the net strangeness number of the total system\cite{CG2}:
\begin{eqnarray}\label{eq1} 
\frac{d}{dt}A^{tot}  & = & -\Gamma_1 \, A^{HG}  \nonumber \\
\frac{d}{dt}S^{tot}  & = & -\Gamma_2 \, S^{HG} \\
\frac{d}{dt}(N_s - N_{\overline{s}})^{tot}  & =  & -\Gamma_3 \,
(N_s - N_{\overline{s}})^{HG} \, \, \, , \nonumber
\end{eqnarray}
For simplicity, we set $\Gamma_1=\Gamma_2=\Gamma_3$,
thus, $\Gamma = \frac{1}{A^{HG}}
\left( \frac{\Delta A^{HG}}{\Delta t} \right) _{ev}$ is the effective
(`universal') rate of particles (of converted hadron gas volume)
evaporated from the hadron phase.
A more general treatment with differing rates for the three quantities is 
presently being studied\cite{Dumi97}.

The equation of state consists of the bag model for the
quark gluon plasma and the hadron gas of the previous section.  
Thus, one solves simultaneously
the `non-equilibrium' dynamics (\ref{eq1}) and the Gibbs phase equilibrium
conditions for the intrinsic variables, i.e. the chemical potentials and the
temperature, as functions of time.

\section{Particle rates from the hadronizing plasma}
\begin{figure}[b]
\vspace*{\fill}
\centerline{\psfig{figure=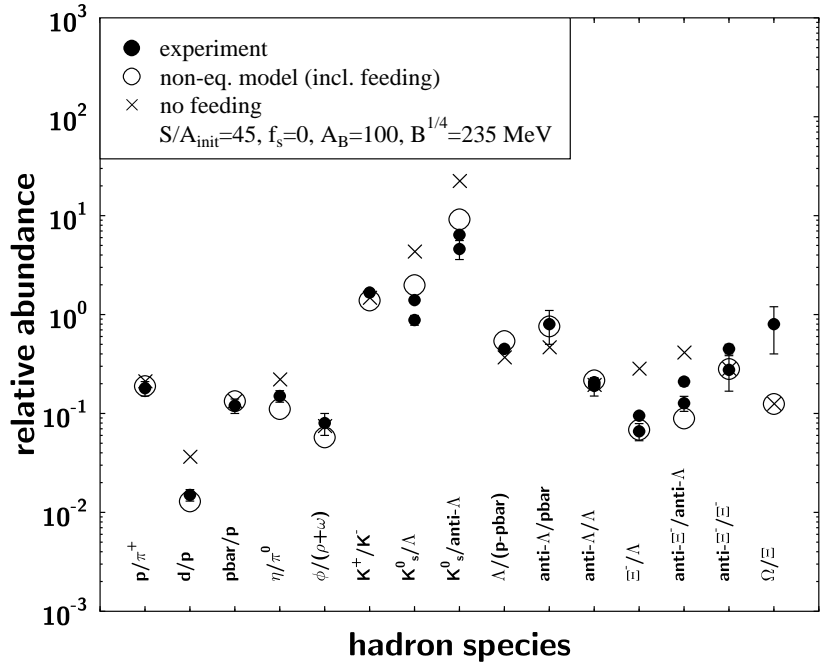,width=12cm}}
\caption{Final particle ratios in the non-equilibrium scenario with initial
conditions $A_{\rm B}^{\rm init}=100$, $S/A^{\rm init}=45$, 
$f_s^{\rm init}=0$ and bag constant $B^{1/4}=235$~MeV. The crosses denote
the resulting values if contributions due to the decay of higher lying
resonances are ignored. The circles include these effects.
Data from various experiments as compiled in 
\protect\cite{PBM} 
are also shown.
\label{ratios45}}
\vspace*{\fill}
\end{figure}

The particle yields as functions
of time have been calculated for different parameter sets: In
Fig.~\ref{ratios45} the final (time integrated)
ratios are plotted for a bag constant of $B^{1/4}=235$~MeV, initial
strangeness fraction $f_s=0$  and an initial
specific entropy per baryon of $S/A=45$. We choose an initial net-baryon 
number $A_B^{init}=100$. Obviously, the particle ratios will not 
depend on this choice.
The theoretical ratios show an
equally good overall agreement with the data points as the static fit.
This scenario results in a rather rapid hadronization. The
quasi-isentropic expansion of the system is due to
those hadrons which are subsequently evaporated. 
Decays and feeding occur after the hadrons
leave the system. 
The specific entropy extracted with this fit is higher than the 
values given in \cite{Ri90} for this beam energy ($S/A\approx 20-25$),
where entropy production was calculated in a one-dimensional hydrodynamic 
model of a heavy ion collision.
The calculation of particle rates from a hadronizing QGP with
$B^{1/4}=235$~MeV has already been presented in \cite{CG2}, however, with
$S/A^{init}=25$. Predictions have been also made for different conditions
within the rate-equation approach of \cite{Ba88} which
yields much higher $K/\pi$ ratios.

\begin{figure}[b]
\vspace*{\fill}
\centerline{\psfig{figure=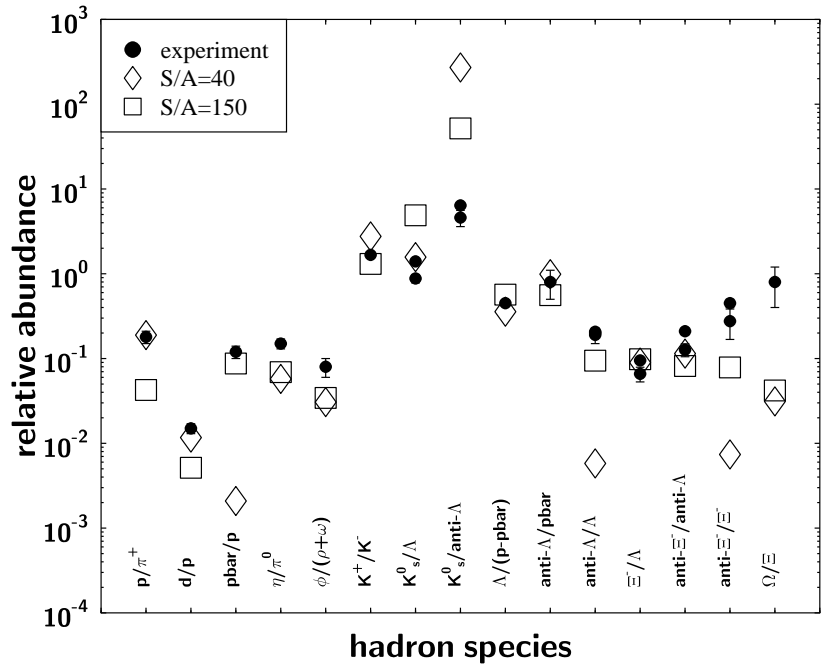,width=12cm}}
\caption{Final particle ratios (including feeding) in the non-equilibrium 
scenario with initial
conditions $A_{\rm B}^{\rm init}=100$, 
$f_s^{\rm init}=0$ and bag constant $B^{1/4}=160$~MeV. 
The initial specific entropy is $S/A^{\rm init}=40$ (diamonds) and
$S/A^{\rm init}=150$ (squares), respectively.
Data from various
experiments as compiled in 
\protect\cite{PBM} 
are also shown.
\label{b160}}
\vspace*{\fill}
\end{figure}

The
system can cool within this model (meaning the employed equations of
state for both phases) if $(S/A)^{QGP}/(S/A)^{had} < 1$
\cite{CG2,Hwa95}, which can only be achieved for a low bag constant.
Low bag constants allow for the formation of metastable strange 
remnants of the
plasma, the strangelets. However, the particle yields as resulting from 
such small
$B$-values in the present model are not in good overall agreement 
with the observed ratios. This is shown in Fig.~\ref{b160} where
a bag constant of $B^{1/4}=160$~MeV and an initial
specific entropy per baryon of $S/A=40$ and $S/A=150$, respectively, is
chosen. The results for the lower specific entropy are tolerable except for
ratios where antibaryon yields are set in relation to baryon or meson
yields. Antibaryon to baryon ratios fail by more than one order of
magnitude, which is due to the lower temperatures of the coexistence phase.
To counteract this effect, much higher initial specific entropies would be
required. However, in particular the ratios $p/\pi^+$ 
and $d/p$ will then move away considerably from the experimentally observed 
values (see Fig.~\ref{b160}). 

Note that within microscopic hadronic model calculations\cite{Ja94} 
a pronounced dependence of antibaryon (and even more so anti-cluster) yields 
on the reaction volume has been predicted. This chemical non-equilibrium
reflects the strong sensitivity of antibaryon production and absorption on 
the phase space evolution of the baryons.
It is in contradiction to the
'volume freeze-out' of thermal models. On the other hand, even if
equilibration of antibaryons is assumed, the abundances are predicted to
be strongly enhanced in the case of an {\em interacting} hadron 
gas\cite{Scha91}.

\begin{figure}[b]
\vspace*{\fill}
\centerline{\psfig{figure=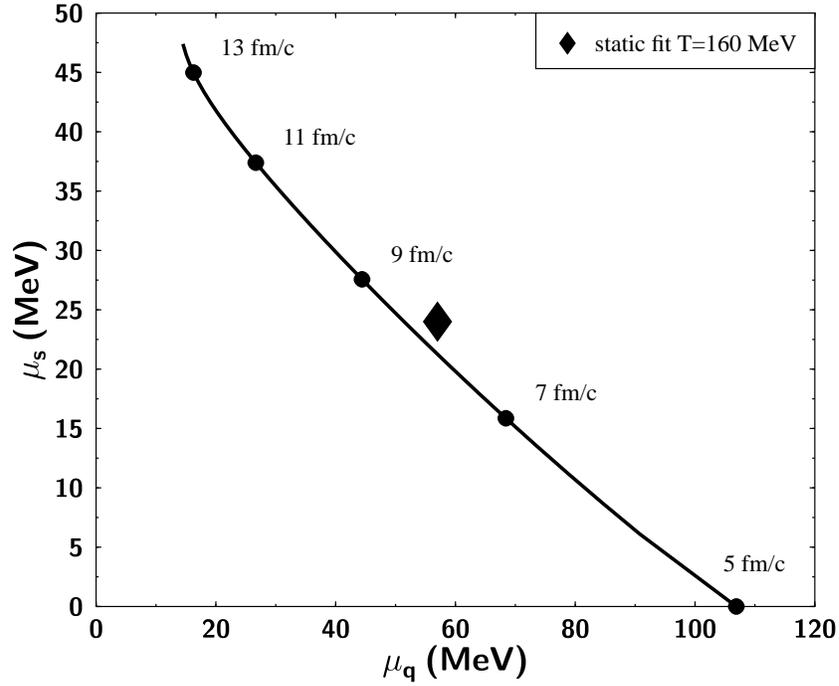,width=12cm}}
\caption{Time evolution of the quark chemical potential and the strange
quark chemical potential for the parameter set of Fig.~\ref{ratios45} 
(here and in the following pictures we start the hadronization at
an initial time of $t_0=5$~fm/c).
The diamond denotes the chemical potentials extracted with the static fit
(Fig.~\ref{fs160}).
\label{mumu}}
\vspace*{\fill}
\end{figure}

Fig.~\ref{mumu} shows the time evolution of the system (initial conditions
as in Fig.~\ref{ratios45}) in the plane  of quark and strange quark chemical
potential.
The quark chemical potential is at the beginning of the evolution in the 
order of the temperature of the system. It is $\mu_q\approx
110$~MeV. This is twice the value
assumed in the static approach, $\mu_q\approx 60$~MeV. 
However, in the course of the
hadronization, the quark chemical potential drops to $\mu_q\approx
15$~MeV. On the other hand, the strange chemical potential increases
from $\mu_s=0$ to
values of $\mu_s\approx 50$~MeV at the end of the evolution, as
compared to $\mu_s\approx 25$~MeV in the static fit. This is due to the
strangeness distillation effect (see below).

Fig.~\ref{muster235} shows the time evolution of the system in the $\mu_q$-$T$
plane in conjunction with the phase diagram of the quark gluon plasma phase
and the hadron gas.
The initial specific entropy of $S/A=45$ in combination with the bag
constant $B^{1/4}=235$~MeV leads to a fast decomposition of the
quark phase within $\approx 13$~fm/c. 
The system moves on a path along the phase boundary.
In our case we met the situation $(S/A)^{QGP}/(S/A)^{HG}>1$ which leads to a
reheating, i.e. $(S/A)^{QGP}$ increases with time\cite{CG2,Sub86,Hwa95}, to
a final value of $S/A\approx 140$. 
As a consequence the
temperature also increases, but only slightly (as can be seen in
Fig.~\ref{muster235}). For $B^{1/4}=160$~MeV the system cools, while the
specific entropy of the quark drop decreases. In this case, the
hadronization is incomplete and a strangelet of baryon number 
$A\approx 8-9$ is formed.

\begin{figure}[b]
\vspace*{\fill}
\centerline{\psfig{figure=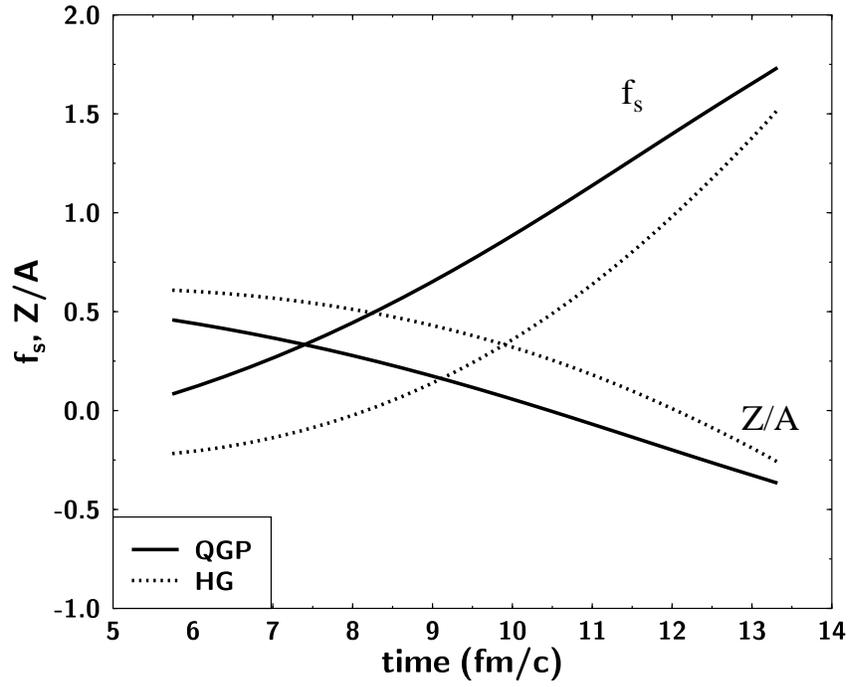,width=12cm}}
\caption{Time evolution of the QGP and the hadron phase 
(initial conditions as in Fig.~\ref{ratios45}).
Shown are the strangeness fraction $f_s^{QGP}$, $f_s^{HG}$ as well as the
fraction of charge over baryon number $Z/A^{QGP}$, $Z/A^{HG}$.
\label{zafs}}
\vspace*{\fill}
\end{figure}

Due to the 'strangeness distillery' effect \cite{CG1}
strange and antistrange quarks do not hadronize
at the same time for a baryon-rich system: Both 
the hadronic and the quark matter phases enter the strange sector,
$f_s\neq 0$, of the phase
diagram immediately. This is valid for both scenarios, the higher bag
constant (reheating) as well as the lower bag constant (cooling).
The effect can be seen from the high values of the
strangeness fraction in the quark phase, shown in Fig.~\ref{zafs} for the 
higher bag constant $B^{1/4}=235$~MeV. At the late stage of the evolution,
as the strange chemical potential increases,
the hadron phase reaches positive $f_s$ values as well. Fig.~\ref{zafs}
shows also the charge to mass ratio of the two phases.

\begin{figure}[b]
\vspace*{\fill}
\centerline{\psfig{figure=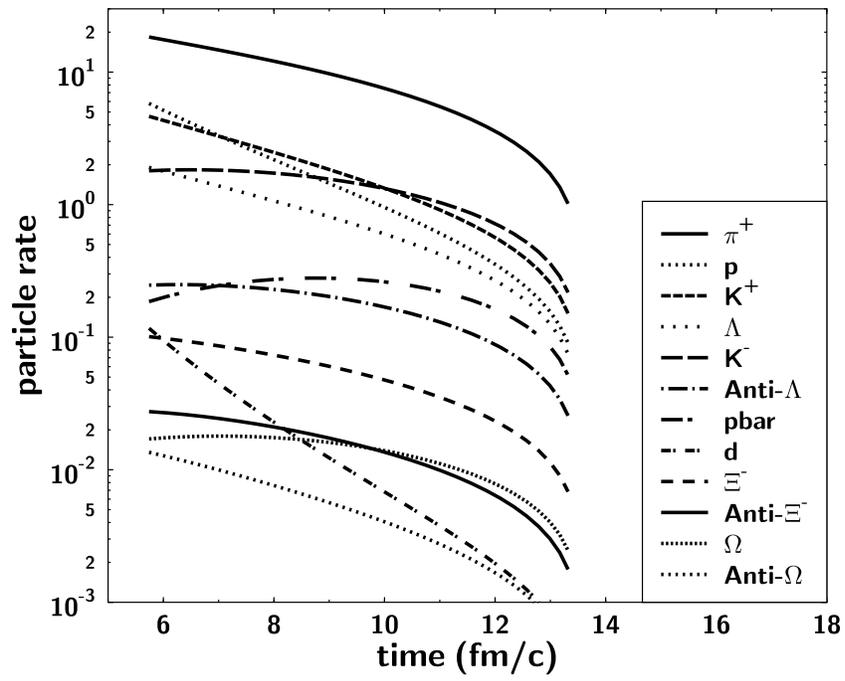,width=12cm}}
\caption{Particle rates as functions of time
(initial conditions as in Fig.~\ref{ratios45}).
\label{trate45}}
\vspace*{\fill}
\end{figure}

The rapid expansion of the system leads to 
 changes of the chemical potentials which reflect on the differential 
hadron production rates. 
Fig.~\ref{trate45} shows the particle rates $dN_i/dt$ for different
hadrons as functions of time for the initial condition $A_B^{init}=100$,
$S/A=45$, $f_s=0$ and a bag constant of $B^{1/4}=235$~MeV (as in
Fig.~\ref{ratios45}).
The particle rates decrease in general
due to the shrinking of the system, since evaporation is proportional to the
surface. However, the differences in the time dependences of different
hadron rates are considerable: proton and deuteron rates drop very fast due
to the decreasing quark chemical potential, while the antiproton rate even
increases (for the same reason). $K^-$ and $\Omega$ production profits 
from the
high strange quark chemical potential at the late stage of the evolution.

\begin{figure}[b]
\vspace*{\fill}
\centerline{\psfig{figure=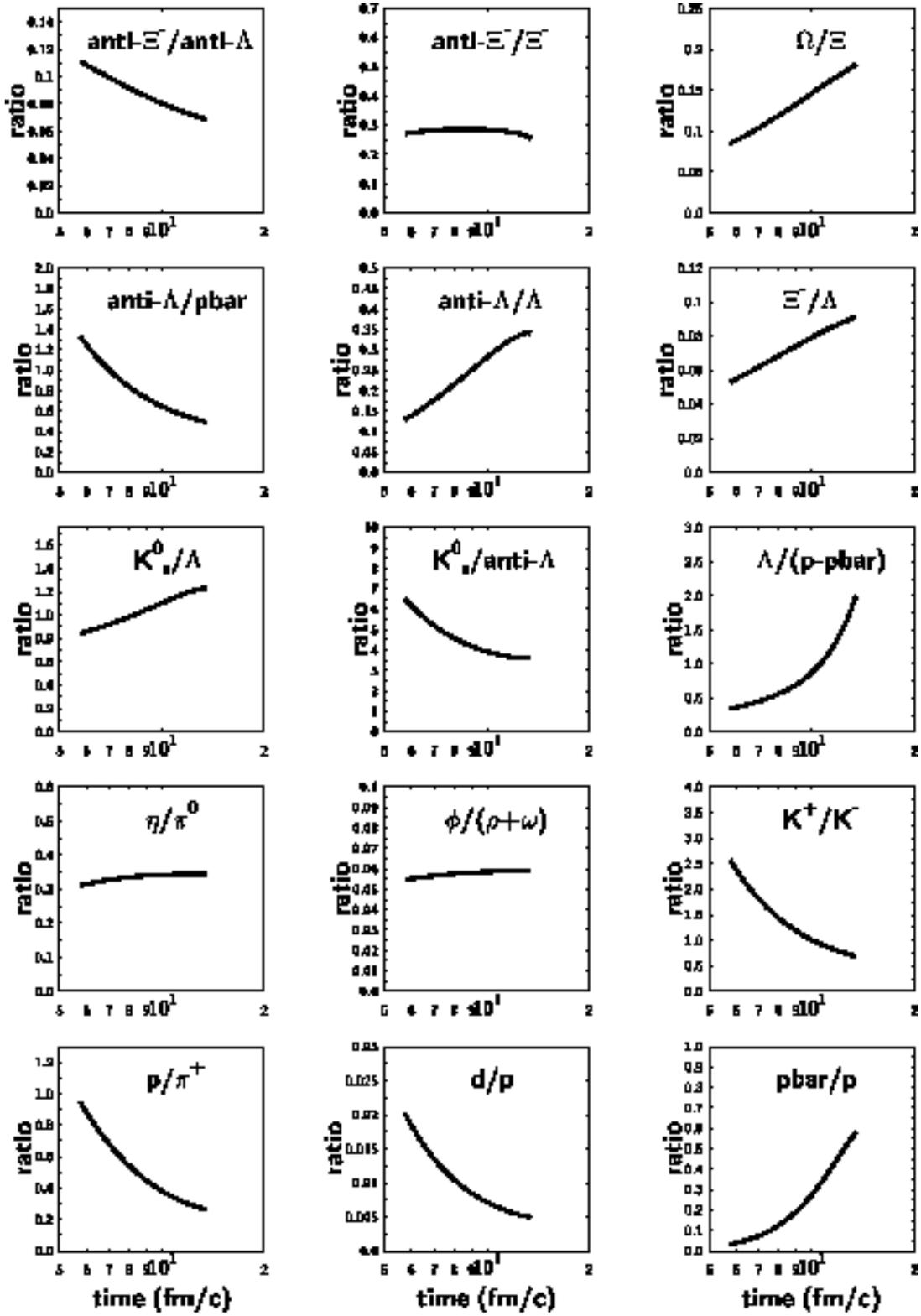,width=15cm}}
\caption{Time evolution of various particle ratios (including feeding)
for the parameter set of Fig.~\ref{ratios45}. 
Here the ratios of the {\em particle
rates} at a certain time $t$ are depicted.
\label{logdmulti}}
\vspace*{\fill}
\end{figure}

Fig.~\ref{logdmulti} shows the time evolution of various particle ratios
for the same calculation.
Most of the ratios -- the final values of
which are given in Figs.~\ref{ratios45} -- change by 
factors $\approx 2-5$ in the course of the hadronization and evaporation. 
Therefore, the thermodynamic parameters of a certain 
stage of the evolution, e.g. of the initial stage, cannot be deduced from
the final integrated particle
ratios in the present model. For the bag constant of $B^{1/4}=235$~MeV the 
temperature stays rather constant
(which is not required by the model). Therefore, ratios which do not depend 
directly 
on the chemical potentials, like $\eta/\pi^0$ and $\phi/(\rho+\omega)$, do 
not change significantly.

Table~\ref{tab3} shows the absolute abundances of different hadron species for
different initial conditions of the hadronizing quark-hadron system. 
We assume an initial baryon number of $A_B=208+208$ (Pb+Pb). However, the
final particle ratios do not depend on this, because the hadrons yields in
this model scale with $A_B$. Thus one can easily re-normalize the model
prediction for any system size, i.e. the number of participant nucleons.
Keep in mind that isospin symmetric systems are assumed.
For $B^{1/4}=235$~MeV the system hadronizes completely, while the specific
entropy of the quark phase rises from $S/A^{init}=30$ ($40$, $50$) to
$S/A^{final}=109$, ($130$, $153$). For
$B^{1/4}=160$~MeV a cold strangelet of mass $A\approx 35$ ($S/A^{init}=40$)
or $A\approx 42$ ($S/A^{init}=150$) emerges. 
The table shows that antibaryon yields are very 
sensitive to the bag constant, whereas the pions depend strongly on the
initial specific entropy.

The $K/\pi$ ratio from the dynamic
hadronization and the 'favorable' bag constant of $B^{1/4}=235$~MeV 
turns out to be too high (this is also a problem of the static fit).
From the preliminary plots in \cite{Jo96} one 
can read off a $K_s^0/h^-$ ratio of $0.11$--$0.12$ for central Pb+Pb
collisions ($4\pi$ yields), whereas the model yields a value of 
$K_s^0/h^-\approx 0.15$ for the three choices of initial specific entropy.
It is possible to improve this ratio by assuming a lower bag constant.
For $B^{1/4}=160$~MeV the model renders $K_s^0/h^-$ ratio of $\approx 0.1$,
as can be seen in Table~\ref{tab3}.
It can be argued that the numbers of kaons and 
pions as the most abundant particles reflect the actual thermodynamic
conditions much more accurately than other hadron species. 

\begin{figure}[b]
\vspace*{\fill}
\centerline{\psfig{figure=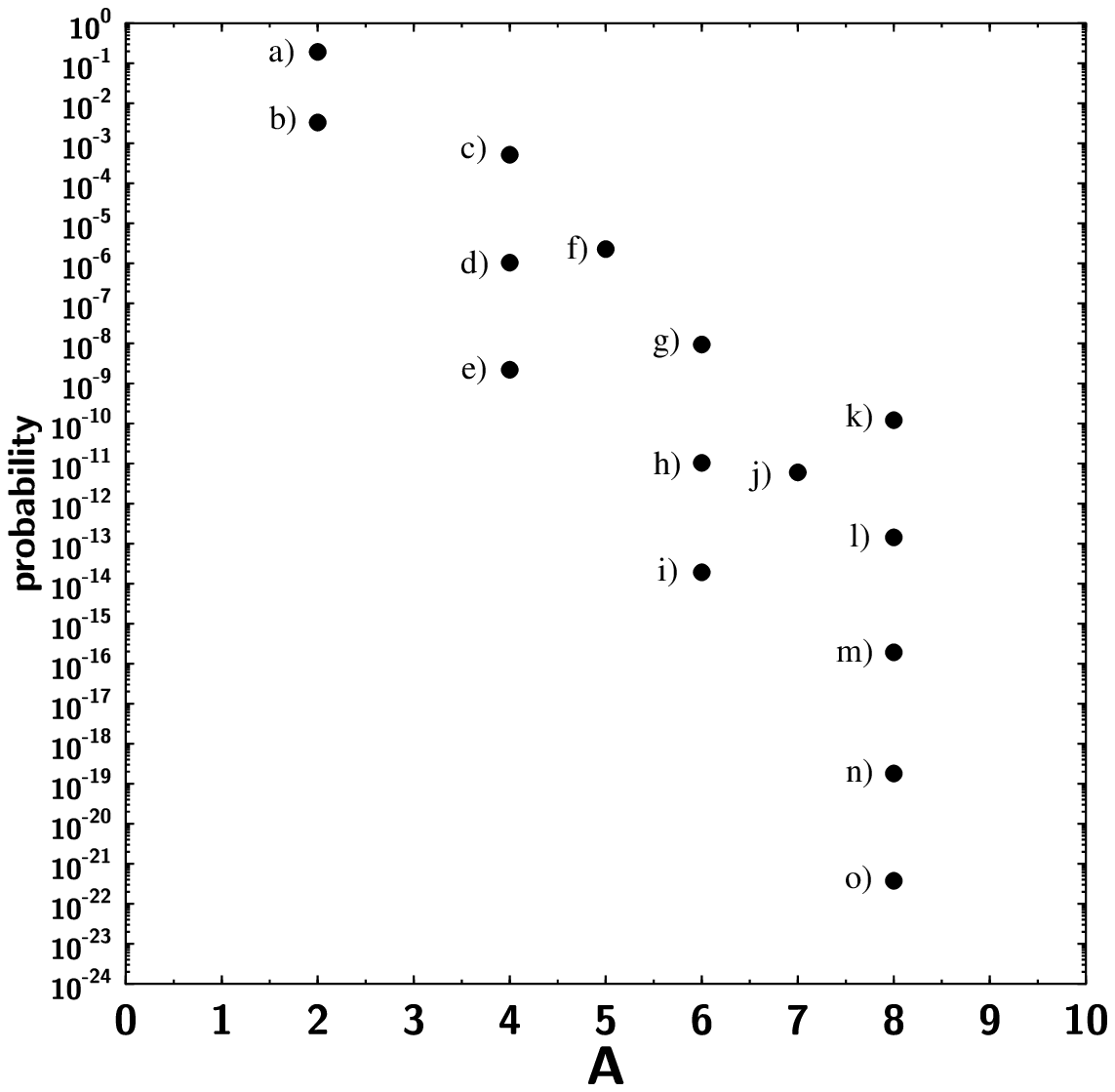,width=12cm}}
\caption{Calculated multiplicities of hypermatter clusters from a 
hadronizing QGP with
$A_B^{init}=416$, $S/A^{init}=40$, $f_s^{init}=0$ and a bag constant of
$B^{1/4}=235$~MeV:\\
a) $H^0$ ($m=2020$~MeV), 
b) \{$\Xi^-,\Xi^0$\},
c) ${}^4He$,
d) \{$4\Lambda$\},
e) \{$2\Xi^-,2\Xi^0$\},
f) ${}^5_\Lambda He$,
g) ${}^{\;\;6}_{\Lambda\Lambda} He$,\\
h) \{$2n,2\Lambda,2\Xi^-$\},
i) \{$2\Lambda,2\Xi^0,2\Xi^-$\},
j) ${}^{\;\;\;\;7}_{\Xi^0\Lambda\Lambda} He$,
k) $A=8$, $S=0$, 
l) $A=8$, $S=-4$, \\
m) $A=8$, $S=-8$, 
n) $A=8$, $S=-12$, 
o) $A=8$, $S=-16$\\
($-S$ gives the number of strange quarks).
\label{hyper}}
\vspace*{\fill}
\end{figure}

Fig.~\ref{hyper} shows calculated multiplicities of various hypermatter
clusters for $B^{1/4}=235$~MeV, $A_B=416$ and $S/A=40$. The penalty factor
$\sim e^{(\mu-m)/T}$ suppresses the abundances of heavy clusters during
the hadronization process, which is reflected by the final yields:
metastable hypermatter can only be produced with a probability $p< 10^{-8}$
for $A \ge 4$ (e.g. a \{$2\Xi^-,2\Xi^0$\} object).

Can one discriminate the static thermal and the hadronizing source
from observations of hadron abundances? Obviously, the final 
particle ratios are nearly identical and therefore not suitable for
discriminating the two scenarios. 
The strong change in the time 
dependent particle rates (Fig.~\ref{logdmulti}) reflects
different average freeze-out times of the particle species. These, in turn,
correspond to different average freeze-out radii, since the quark drop shrinks
--- in this scenario --- during the hadronization process. Both quantities,
which characterize the size and the lifetime of the particle emitting source,
are in principle accessible by means of Hanbury-Brown-Twiss-analyses\cite{HBT56}. 
This concept is used
extensively for high energy heavy ion collisions (for a review see
\cite{Jac95}). 

In Table~\ref{tab2} the average freeze-out times of
different hadron species resulting from a hadronizing QGP drop 
(for different initial conditions)
are listed. 
Only the directly produced hadrons are taken into account. The
true freeze-out times are generally larger due to the finite lifetime of the
resonances, which contribute to specific hadron yields. However, the
two-particle correlations within the source --- imposed by the quantum
statistical momentum distributions --- are lost through these decays. Thus,
the measured correlation function does not represent a simple Fourier-like 
transform
of the particle distribution inside the emitting source, if the
contributions from resonance decays are dominant. As can be seen from
Table~\ref{tab2} only a fraction of the integrated, final yield of protons, 
kaons and pions are directly emitted from the hadronizing system.
A major part stems from the decay of higher resonances.
 This makes
the accessibility of the freeze-out radii or times by means of HBT analysis
questionable. However, the presented results suggest
that the observed freeze-out time will be smaller
for $K^+$ than for
$K^-$, and it will also be smaller for $p$ and $\Lambda$ than for $\bar p$ and
$\bar \Lambda$, respectively. The corresponding radii behave the opposite in
this model, because the system shrinks.

\section{Conclusion}
Experimental particle ratios at SPS energies are compatible with the
scenario of a static thermal source in chemical equilibrium 
\cite{PBM}. A simple
dynamic hadronization model, however, reproduces the numbers equally well.
It yields the following conclusions: 
a high bag constant of $B^{1/4}\ge 200$ is mandatory. These
values of $B$ would exclude the existence of stable, cold strangelets, 
as hadronization proceeds without any cooling but with reheating.
In central collisions of S+Au(W,Pb) a specific entropy per baryon of
$S/A=35-45$ is created. 
Due to strangeness distillation the system moves rapidly
out of the $T, \mu_B$ plane, into the $\mu_s$-sector.
The quark chemical potential drops during the
evolution, the strange quark chemical potential rises. Final $\mu_q$,
$\mu_s$ values of $1/3$ resp. $3$ of the
values of the static fits are reached for $t\rightarrow t_{freeze-out}$.
We have
presented the model calculations of absolute yields of different hadron species 
for Pb+Pb collisions at SPS energies.
The average freeze-out times of different hadron species differ strongly,
which might be observable via HBT-analyses. Strangeness to baryon fractions
of $f_s\approx 1-2$ suggest that '$\Lambda$-droplets' or even
'$\Xi^-$-droplets' form the system at the late stage.

\begin{table}[h]  
\caption{Hadron abundances for Pb(160GeV/u)Pb ($A_B^{init}=416$, $f_s=0$) and
different initial conditions of the hadronizing quark drop. The numbers for
the total yields (including feeding from higher resonances) are given.
}
\label{tab3}
\begin{tabular}{|cc|cccccccc|}
 $S/A^{init}$ & $B^{1/4}$~(MeV)& $p$ & $\pi^+$ & $d$ & $\bar p$ & $\phi$ & $\rho$& $\omega$ & $K^+$\\ \hline 
 30 & 235&173.7&619.3&4.3&11.3&12.9&176.6&56.0&130.0\\ 
 40 & 235 &180.0&852.2&2.8&20.0&18.8&248.9&81.2&175.4\\ 
 50 & 235 &188.7&1084.5&2.2&30.2&24.6&320.7&106.2&219.2\\ 
 40 & 160 &150.7&798.9&1.8&0.3&2.7  &68.9 &20.8 &124.6\\ 
 150 & 160 &149.7&3522.0 &0.8&13.1&17.6&390.5&122.2&430.3\\ \hline
    &     & $K^-$ & $K^0_s$ & $\Lambda$ & $\bar \Lambda$ & $\Xi^-$ & $\bar \Xi^-$& $\Omega$ & $\bar \Omega$\\ \hline 
 30 & 235 &81.8&105.9&67.6&8.6&4.4&0.81&1.00&0.30\\ 
 40 & 235 &122.3&148.8&79.8&15.0&5.4&1.3&1.3&0.49\\ 
 50 & 235 &163.4&191.3&90.5&22.5&6.4&2.1&1.6&0.70\\ 
 40 & 160 &44.8&84.7&53.8&0.3&4.9&0.04&0.3&0.01\\ 
 150 & 160 &328.7&379.5&77.6&7.4&7.7&0.6&0.6&0.01\\ 
\end{tabular}
\end{table}

\begin{table}[h]  
\caption{Average freeze-out times (in fm/c) of different hadron species for 
Pb(160GeV/u)Pb ($A_B^{init}=416$, $f_s^{init}=0$).
The results for different initial specific entropies of the hadronizing 
quark drop and different bag constants are shown.
The fraction of directly produced hadrons divided by the total
yield (including feeding) is given in brackets.
}
\label{tab2}
\begin{tabular}{|cc|cccccc|}
 $S/A^{init}$ & $B^{1/4}$~MeV  & $p$ & $\pi^+$ & $d$ & $\bar p$     & $K^+$      & $K^-$      \\ \hline 
30& 235 & 3.3 (.36) & 4.8 (.30) &2.1 (1.0) & 7.1 (.36)             & 4.2 (.55)  &6.0 (.51) \\  
40& 235 & 4.1 (.36) & 5.6 (.32) &2.9 (1.0) & 7.6 (.36)            & 5.0 (.54) &6.6 (.51) \\  
50& 235 & 4.8 (.35) & 6.2 (.32)   &3.5 (1.0)& 8.1 (.35)           & 5.6 (.53) &7.1 (.51) \\ \hline
  &        & $\Lambda$     & $\bar \Lambda$      & $\Xi^-$      & $\bar \Xi^-$      &  $\Omega$ & $\bar \Omega$\\ \hline 
30& 235    & 4.3 (.24)     & 5.8 (.21)          & 5.1 (.98)   & 5.2 (.97)       &6.1 (1.0)    & 4.3 (1.0)\\  
40& 235    & 5.1  (.24)     & 6.4 (.21)          &5.8 (.98)   & 5.8 (.97)       &6.8 (1.0)    & 5.0 (1.0)\\  
50& 235    & 5.8 (.24)     & 6.9 (.22)          & 6.4 (.98)   & 6.4 (.97)       &7.3 (1.0)    & 5.6 (1.0)\\ \hline
\end{tabular}
\end{table}

\acknowledgements
We thank Dieter R\"ohrich, Reinhard Stock, Johanna Stachel, Peter
Braun-Munzinger, Johannes Wessels and Helmut Satz for 
fruitful discussions.

\clearpage
\clearpage

\clearpage

\clearpage

\clearpage

\begin{figure}[b]
\vspace*{\fill}
\centerline{\psfig{figure=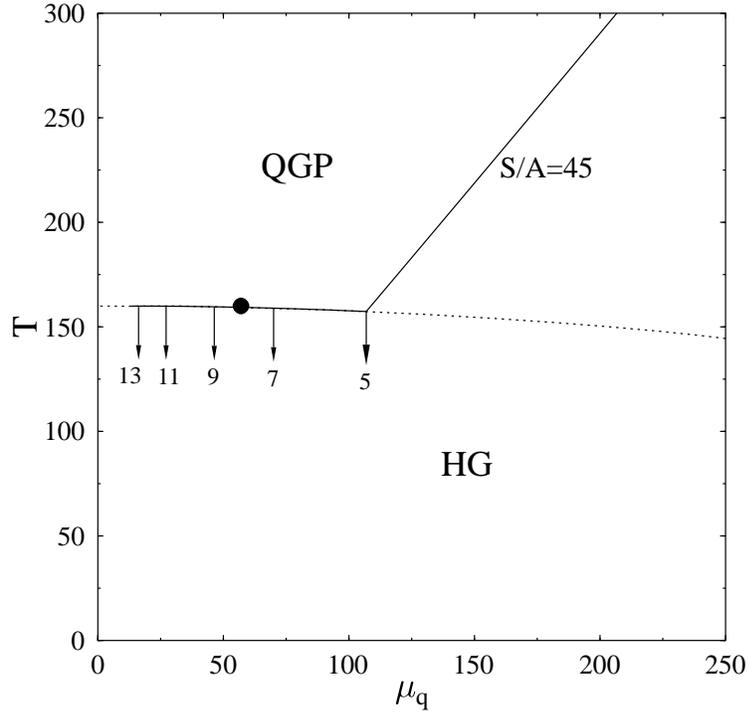,width=12cm}}
\caption{Hadronization path of the system (initial conditions as in
Fig.~\ref{ratios45}) in the projected $T-\mu_q$ plane. The numbers denote 
the 
time in fm/c.
Shown is also the phase
boundary of a QGP and the hadron gas for $B^{1/4}=235$~MeV and the path of
constant specific entropy in a pure quark phase with $S/A=45$. The full
circle indicates the $\mu_q$ and $T$ values of the static fit.
\label{muster235}}
\vspace*{\fill}
\end{figure}
\clearpage

\clearpage

\clearpage

\clearpage

\end{document}